\documentclass[fleqn,usenatbib]{mnras}

\usepackage{newtxtext,newtxmath}

\usepackage[T1]{fontenc}

\DeclareRobustCommand{\VAN}[3]{#2}
\let\VANthebibliography\thebibliography
\def\thebibliography{\DeclareRobustCommand{\VAN}[3]{##3}\VANthebibliography}

\usepackage{graphicx}	
\usepackage{amsmath}	

\title[On the resolution of eclipse mapping]{An analytic theory for the resolution attainable using eclipse mapping of exoplanets}

\author[Boone, Grant, \& Hammond]{
Sasha Boone$^{1}$\thanks{E-mail: alexander.boone@hertford.ox.ac.uk},
David Grant$^{2}$,
\&
Mark Hammond$^{1}$
\\
$^{1}$Department of Physics, University of Oxford, Oxford, UK\\
$^{2}$University of Bristol, HH Wills Physics Laboratory, Tyndall Avenue, Bristol, UK\\
}

\date{Accepted XXX. Received YYY; in original form ZZZ}

\pubyear{2023}

\begin{document}
\label{firstpage}
\pagerange{\pageref{firstpage}--\pageref{lastpage}}
\maketitle

\begin{abstract}
We present an analytic theory for the resolution attainable via eclipse mapping of exoplanets, based on the Fourier components of the brightness distribution on the planetary disk. We find that the impact parameter determines which features can and cannot be seen, via the angle of the stellar edge relative to the axis of the orbit during the eclipse. We estimate the signal-to-noise ratio as a function of mapping resolution, and use this to determine the attainable resolution for a given star-planet system. We test this theory against numerical simulations and find good agreement; in particular, our predictions for the resolution as a function of stellar edge angle are accurate to the simulated data to within 10\% over a wide range of angles. Our prediction for the number of spatial modes that can be constrained given a light curve error is similarly accurate. Finally, we give a list of exoplanets with the best expected resolution for observations with the NIRISS SOSS, NIRSpec G395H, and MIRI LRS instruments on JWST. 
\end{abstract}





\section{Introduction}
Understanding the temperature distributions of exoplanets is critical to testing models of heat redistribution, which in turn grant valuable insight into the mechanisms driving heat transport and the dynamic structure of the atmosphere \citep{lewis2022temperature}. While simulations and theoretical calculations (e.g. \citet{perez2013atmospheric}) can inform us about what the temperature distributions of exoplanetary atmospheres might look like, these results cannot be tested without observational data.

Phase curves, which are observations of phase-resolved emission from planets, provide longitudinal information about their brightness temperatures. Their day-side and night-side temperatures, and the offsets of their hottest points from the substellar point, are especially valuable information. By decomposing the phase curve into its Fourier components, the longitudinal brightness distribution of the planet can be determined \citep{cowan2008inverting}; this method was successfully applied to HD 189733 b \citep{knutson2007map}. Planets with axes of rotation inclined to the line of sight provide very limited latitudinal information, in that the spherical harmonics antisymmetric about the equator will also contribute to the light-curve \citep{cowan2013light}. 

Eclipse mapping, a method first used on stars and accretion disks \citep{horne1985images, baptista2004eclipse, collier1997eclipse}, is now available as a means to study exoplanets thanks to the precision of measurements with the James Webb Space Telescope (JWST) \citep{rauscher2007toward}. As a planet is eclipsed by its star, different parts of the planet are hidden at different times. Measuring the light curve therefore provides information about the brightness distribution of the planet \citep{williams2006resolving, majeau2012two}. Eclipse mapping allows asymmetries about the equator to be constrained (as seen in simulations in e.g. \citet{menou2020hot}, \citet{cho2003changing}), and since secondary eclipses are short compared to the planet's orbital period, the method takes less telescope time and assumes less about the stability of brightness features on the planetary photosphere \citep{jackson2019variability, rauscher2007hot, cho2003changing}. 

While eclipse mapping was attempted by \citet{majeau2012two,de2012towards} using Spitzer data of HD 189733 b, they could only constrain the spherical harmonics up to $l=1$, i.e. the dipole moment, and three of the four modes could be constrained with only the phase curve. Using JWST, it is now possible to apply eclipse mapping to some exoplanets and achieve reasonable resolution on the day side of the planet using a relatively short amount of telescope time; \citet{coulombe2023broadband} were able to constrain five spherical harmonics on WASP-18 b using a segment spanning only around a third of the full orbit, although there was a low degree of confidence in the fitted latitudinal structure. In general, eclipse mapping results cannot be trusted to very high resolution due to the presence of a ``null space'' of spatial patterns that do not contribute to a given eclipse signal \citep{challener2023eclipse}.

Hot Jupiters are gas giants in extremely short orbits around their stars ($P < \sim 10$ days).  They have high equilibrium temperatures because of their proximity to their host stars, and are often inflated with radii moderately larger than Jupiter's. Their high temperatures and large sizes give them excellent signal-to-noise ratios (SNRs) and thus make them ideal targets for mapping. They are expected to be tidally locked due to their close-in orbits \citep{rasio1996tidal, guillot1996giant}, and global circulation model (GCM) simulations predict them to have large day-night temperature contrasts due to the strong stellar heating \citep{showman2002atmospheric, perez2013atmospheric}. Simulations also find a strong superrotating component at the equator, which in turn results in a shift of the hot-spot eastwards and a warming of the equator towards the terminator \citep{menou2020hot}. These features are validated by observations, with large phase curve amplitudes and phase offsets being ubiquitous among hot Jupiters \citep{parmentier2017exoplanet, knutson2007map}, though GCMs often have difficulty replicating the observed values \citep{showman2020atmospheric}. 

The inclusion of drag (often in the form of Lorentz drag by the planetary magnetic field on a partially ionised atmosphere \citep{menou2012magnetic}) can serve to hinder heat transport and thus results in a sharper drop in temperature at the terminator \citep{perez2013atmospheric}. Clouds may also form on the night side and the cooler western hemisphere of hot Jupiters \citep{parmentier20133d}, and thereby create a larger thermal contrast and a reversed optical phase curve offset compared to the thermal phase curve \citep{roman2021clouds}. Clouds, photochemistry, and other processes may also create chemical gradients which will affect the temperature profile and spectrum emitted by the atmosphere \citep{parmentier2018thermal}. 

While phase curves can probe some of these, eclipse mapping allows hot Jupiters to be observed in greater detail, and can access features which the phase curve alone cannot. It is clear that the resolution of an eclipse map can be improved by observing targets with better SNR and longer eclipses, and it might seem obvious that a planet with a perfectly edge-on orbit will not be able to provide much latitude information due to the system being symmetric along that axis. One might also guess that eclipse observations might be more sensitive to sharp gradients (e.g. as found in \citet{parmentier2016transitions}) than the phase curve, as the phase curve only gives hemisphere-averaged flux. That said, it is not immediately obvious how various parameters of the planet and its orbit affect the resulting eclipse map. We therefore derive a quantitative prediction of the expected resolution of an eclipse map derived from an observation of any planet given its properties, its orbital parameters, and the precision of the observation.

In Section \ref{sec:theory}, we provide an argument for the amount of degeneracy in eclipse mapping. We then derive the light-curve amplitude for sinusoidal surface modes, and finally derive the resolution achievable using eclipse mapping for surface modes up to an order $N$. In Section \ref{sec:methods}, we discuss the methods used in our numerical tests. In Section \ref{sec:results}, we present the results of those numerical tests. In Section \ref{sec:real_planets}, we apply our theory to real planets to predict which make the best targets for eclipse mapping. Finally, in Section \ref{sec:conclusions}, we give our conclusions.

\section{Theory}\label{sec:theory}
\begin{figure*}
    \centering
    \includegraphics[width=\textwidth]{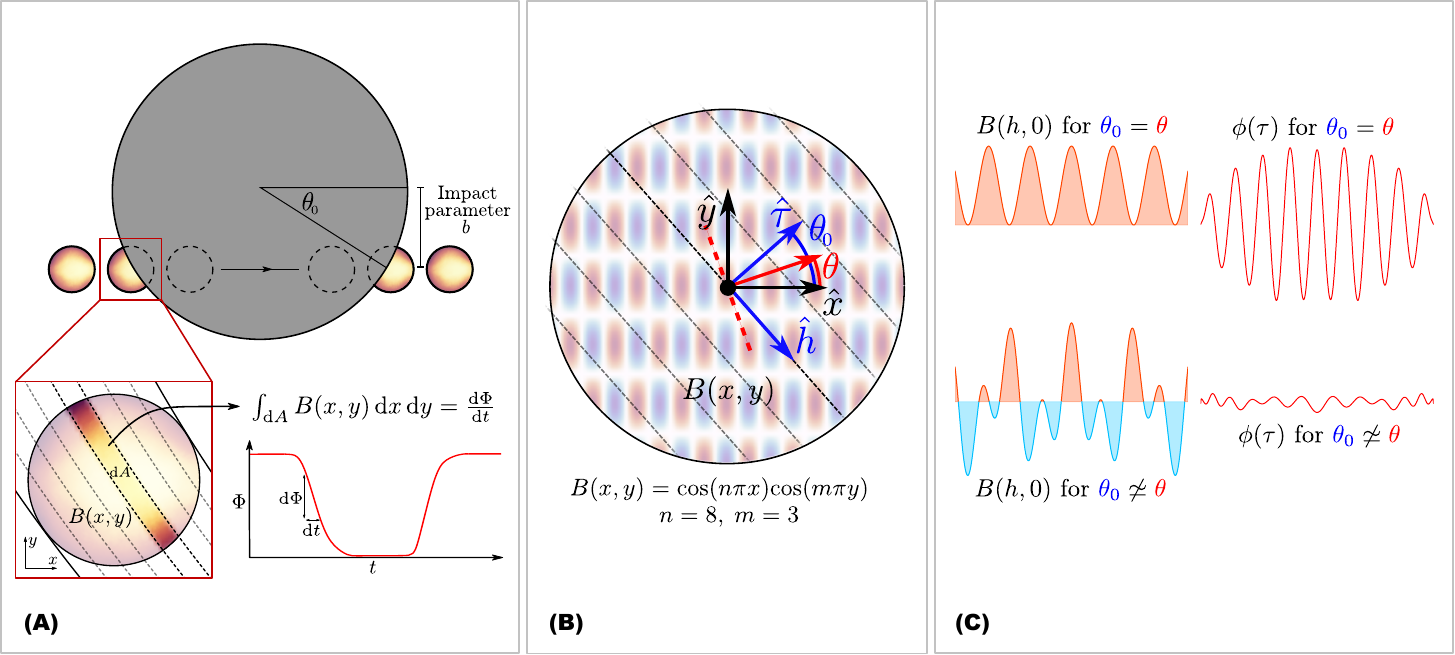}
    \caption{\textbf{(A)} The premise of eclipse mapping. As a planet passes behind its star, different slices of the planet are hidden at different times. The change in sign of the stellar edge angle from ingress to egress (and to a lesser extent, the curvature of the star) gives 2D information about the brightness distribution. During some time $\mathrm{d}t$, an area of the planet $\mathrm{d}A$ is hidden. The flux contributed by $\mathrm{d}A$ is $\mathrm{d}\Phi$, equal to the integral of the brightness distribution $B(x,y)$ over the slice. \textbf{(B)} The geometry we use in our derivation. We use Cartesian coordinates rather than spherical ones as they lend themselves more easily to analytic calculations. $h$ is a coordinate defined along the stellar edge, while $\tau$ is a modified time coordinate perpendicular to the stellar edge. We also show the angle of the stellar edge $\theta_0$ and the angle associated with the mode $\theta$. The dashed red line shows the stellar edge for $\theta_0 = \theta$; this clearly lines up with the features of $B(x,y)$. \textbf{(C)} The brightness contributions from points along the stellar edge for the mode shown in (B), when it is well-aligned (top) and poorly-aligned (bottom; see Equation \ref{eq:aligned}). For simplicity, we rotate the coordinates from $B(x,y)$ to $B(h, \tau)$. The corresponding differential light curves $\phi(\tau)$ are also shown; the cancellation between positive and negative sections of $B(h,0)$ when the mode is poorly-aligned result in a very small differential light-curve compared to when the mode is well-aligned.}
    \label{fig:eclipsemapping}
\end{figure*}
Eclipse mapping works by measuring the brightness of a system as a planet is eclipsed by its star. Since different parts of the planet get eclipsed at different points in time, this yields some information about the brightness distribution on the surface of the planet (see Figure \ref{fig:eclipsemapping}A).

In this paper, we are looking to study eclipse mapping with a simple, approximate, analytic model; while this involves presenting a theory for converting from the light curve to the brightness distribution and a numerical method in Section \ref{sec:methods}, we do not seek to replace the existing methods, merely to explain their properties. We would like to refer readers to formalisms such as Starry \citep{luger2019starry} and SPIDERMAN \citep{louden2018spiderman} for actually retrieving brightness distributions from light-curves.

We are looking to determine the level of degeneracy in eclipse mapping, and for an estimate of the strength of the contribution of a given surface mode (e.g. a spherical harmonic or a Fourier mode) to the eclipse light curve. We begin by defining the differential light curve $\phi$:
\begin{equation}
    \phi = \frac{\mathrm{d}\Phi}{\mathrm{d}\tau},
\end{equation}
where $\Phi$ is the light curve and $\tau$ is a modified time coordinate such that $\tau$ ranges from $-1$ to $1$ as the planet goes from fully eclipsed to fully visible, with $\tau=0$ halfway between first and second (or third and fourth) contacts. The differential light curve is easy to calculate assuming $R_p << R_s$ (where $R_p$ and $R_s$ are the planetary and stellar radii respectively) as the integral along a line across the planet (the line being the edge of the star at time $\tau$). We neglect the rotation of the planet between ingress and egress for simplicity. We also assume that $1-b >> R_p/R_s$ so that the angle of the stellar edge does not change over the course of the ingress/egress, which means grazing eclipses are not covered by our theory. Finally, we assume a known ephemeris; we do not include any errors in the eclipse timing. While the hotspot offset is degenerate with the timing of the eclipse (predicted by \citet{williams2006resolving} and observed by \citet{agol2010climate}), measurements made with JWST are precise enough to break this degeneracy; \citet{coulombe2023broadband} were able to constrain eclipse timing to within 9s for an ingress/egress duration of over 15 minutes. Spectrally resolved observations can also break this degeneracy, as maps can vary as a function of wavelength, but system parameters cannot \citep{dobbs2015spectral}.

\subsection{Some geometry}
We define an angle $\theta_0$ such that the impact parameter $b = \mathrm{sin}\theta_0$ and thus $\sqrt{1-b^2}=\mathrm{cos}\theta_0$. We take the planet to be a circle of radius 1 centred on the origin, and call the coordinates parallel and perpendicular to the direction of motion $x$ and $y$ respectively. We find the stellar edge forms a line such that:
\begin{align}
     x = \tau\,\mathrm{cos}\theta_0 - h\,\mathrm{sin}\theta_0 \\
     y = \tau\,\mathrm{sin}\theta_0 + h\,\mathrm{cos}\theta_0       
\end{align}

where $h$ is a coordinate parallel to the stellar edge, and can take values such that $|h| < \sqrt{1-\tau^2}$ (see Figure \ref{fig:eclipsemapping}B). For some brightness distribution $B(x,y)$, the differential light-curve is:
\begin{align}\label{eq:difflc}
    \phi(\tau) &= \int_{-\sqrt{1-\tau^2}}^{\sqrt{1-\tau^2}} B(x,y)\,dh \\
    &= \int_{-\sqrt{1-\tau^2}}^{\sqrt{1-\tau^2}}B(\tau\,\mathrm{cos}\theta_0-h\,\mathrm{sin}\theta_0, \tau\,\mathrm{sin}\theta_0+h\,\mathrm{cos}\theta_0)\,\mathrm{d}h.
\end{align}

\subsection{The degeneracy} 
Let us take a differential light-curve of the form
\begin{equation}
    \phi(\tau) = \sqrt{1-\tau^2}\, P_N(\tau)
\end{equation}
where $P_N$ is a polynomial of order $N$. Consider a separable brightness pattern produced from the product of a polynomial of order $n$ in $x$, and another of order $m$ in $y$:
\begin{equation}
    B(x,y) = p_n(x)\, q_m(y),
\end{equation}
which produces the differential light-curve $\phi$. The result of this product will be some order $n+m$ polynomial in $h$ and $\tau$. Examining the contribution to the light curve of one term of this polynomial,
\begin{equation}
    \phi_{ij} = \int_{-\sqrt{1-\tau^2}}^{\sqrt{1-\tau^2}} h^i\, \tau^{j} dh =       
    \begin{cases}
        \sqrt{1-\tau^2}\,\frac{2}{i+1}(1-\tau^2)^{\frac{i}{2}}\,\tau^{j} & i\,\, \text{is even}\\
        0 & i\,\, \text{is odd}
    \end{cases}
\end{equation}
and thus contributes a polynomial of order $i+j$ to $P_N$. As $i + j = n + m$, we must set $n + m = N$ if we want the light curve produced by $B$ to be a polynomial of the same order as $\phi$. For $B$ to regenerate the differential light curve $\phi$, we must match the coefficients of $p$ and $q$ such that they equal the coefficients on $P_N$. When both $n$ and $m$ are non-zero, the resulting equations are non-linear and involve products of the coefficients of $p$ and $q$. Nonetheless, we have $N$ equations and $n+m=N$ coefficients to solve for, and so can reduce $p,q$ to at most a finite number of solutions.

While for a given $n,m$, we can solve for the coefficients of $p$ and $q$, we must recall that we do not know \textit{a priori} which values of $m$ and $n$ to take, and so \textit{any} $n,m$ pair can satisfy the equations as long as $n+m=N$: we instead have $\frac{1}{2}(N+2)(N+1)$ coefficients to solve for (the number of terms $x^iy^j$ where $i+j\leq N$). Looking closely, we find another $N$ equations given to us by the fact that the ingress and egress light curves will in general be different, and in turn if $p_n$ is odd, the ingress and egress light curves differ by a sign flip, while if $p_n$ is even, they are identical. As there is one $N=0$ mode, we must add that too. As such, we have $\frac{1}{2}(N+2)(N+1)$ coefficients to solve for with $2N+1$ equations; for $N\geq 2$, there is no unique solution. Equivalently, by subtracting solutions from one another, we get a set of brightness distributions which yield light curves equal to zero - the `null space' \citep{challener2023eclipse}. While we cannot determine the true surface brightness distribution from the light curve due to these degeneracies, we can nonetheless make progress using some reasonable assumptions.

\subsection{Sinusoidal brightness patterns}
To examine the observable brightness of the various modes, we will take surface modes of the form:
\begin{equation}
    B_{nm}(x,y) = \mathrm{sin}(n\pi (x-x_0))\,\,\mathrm{sin}(m\pi (y-y_0))
\end{equation}
where $n$ and $m$ are the horizontal and vertical wavenumbers; we can take the overall order $N$ such that $N^2 = m^2 + n^2$ (for reasons that will become clear later on), though the exact combination is arbitrary. We include a phase factor along $x$ and $y$ for generality, though it should not have any effect on the ultimate result. This separable form results in analogous equations between the ingress and egress light curves, so we will only have to treat one and get the other for free (with a sign change in $\mathrm{cos}\theta_0$).

\subsection{Calculating the differential phase curve}
From Equation \ref{eq:difflc}, the differential light curve can be calculated as: 
\begin{align}
    \phi(\tau) = \int_{-\sqrt{1-\tau^2}}^{\sqrt{1-\tau^2}} &\mathrm{sin}(n\pi (x-x_0))\,\,\mathrm{sin}(m\pi (y-y_0))\,\mathrm{d}h\\
    = \int_{-\sqrt{1-\tau^2}}^{\sqrt{1-\tau^2}} &\mathrm{sin}(n\pi (\tau\,\mathrm{cos}\theta_0-h\,\mathrm{sin}\theta_0-x_0))\notag\\
    \times\,&\mathrm{sin}(m\pi (\tau\,\mathrm{sin}\theta_0+h\,\mathrm{cos}\theta_0-y_0))\,\mathrm{d}h.
\end{align}
Using the identity
\begin{align}
    &\int \mathrm{sin}(ax + b)\,\, \mathrm{sin}(cx + d) \,\mathrm{d}x \notag\\
    &= \frac{1}{2}\left(\frac{\mathrm{sin}((a-c)x + (b-d))}{a-c}-\frac{\mathrm{sin}((a+c)x+(b-d))}{a+c}\right),
\end{align}
and collecting like terms in $h$ and $\tau$, we find
\begin{align}\label{eq:twofreq}
            \phi(\tau)=&-\frac{1}{2} \Bigg[\mathrm{sin}\bigg(\pi ((n\,\mathrm{cos}\theta_0 - m\,\mathrm{sin}\theta_0)\,\tau+(-n\,\mathrm{sin}\theta_0-m\,\mathrm{cos}\theta_0)\,h\notag\\
            &+\,(-n\,x_0+m\,y_0))\bigg) (\pi(n\,\mathrm{sin}\theta_0 + m\,\mathrm{cos}\theta_0))^{-1}\notag\\
    &+\,\mathrm{sin}\bigg(\pi((n\,\mathrm{cos}\theta_0+m\,\mathrm{sin}\theta_0)\,\tau +(-n\,\mathrm{sin}\theta_0+m\,\mathrm{cos}\theta_0)\,h\notag\\
        &+\,(-n\,x_0-m\,y_0))\bigg)(\pi(-n\,\mathrm{sin}\theta_0 + m\,\mathrm{cos}\theta_0))^{-1}\Bigg]_{-\sqrt{1-\tau^2}}^{\sqrt{1-\tau^2}}.
\end{align}
We will discard the first term for two reasons: firstly, as $h$ is slowly-varying around $\tau=0$, the dominant frequency contributed by the term is $\sim \pi(n\,\mathrm{cos}\theta_0 - m\,\mathrm{sin}\theta_0)$, which will end up being degenerate with the lower-order terms (which we expect to both be intrinsically brighter and to be damped by a much smaller denominator). Secondly, its denominator is always of order $N$, whereas the second term has a denominator which vanishes for a specific $(n,m)$ pair (which we will call `well-aligned'). At this well-aligned $(n,m)$ pair, the the bright and dark areas of $B(x,y)$ line up with the stellar edge, and so high-frequency (orthogonal) component of the resulting signal will be large (see Figure \ref{fig:eclipsemapping}C, where $\theta \simeq \theta_0$ occurs when the mode becomes well-aligned). Note that neglecting this term does require us to assume that neither of $(n,m)$ is zero (i.e. that the mode varies along both axes); otherwise, the two terms are the same to within a phase difference. Thus we are left with:
\begin{equation}
    \begin{split}
        \phi(\tau)\simeq- \frac{1}{2}\Bigg[\mathrm{sin}\bigg(\pi((n\,\mathrm{cos}\theta_0+m\,\mathrm{sin}\theta_0)\,\tau+(-n\,\mathrm{sin}\theta_0+m\,\mathrm{cos}\theta_0)\,h\\
        -(n\,x_0+m\,y_0))\bigg)(\pi(-n\,\mathrm{sin}\theta_0 + m\,\mathrm{cos}\theta_0))^{-1} \Bigg]_{-\sqrt{1-\tau^2}}^{\sqrt{1-\tau^2}}.
    \end{split}
\end{equation}
Expanding this using the angle summation formula yields
\begin{align}
    \phi(\tau) \simeq -\frac{1}{2}\Bigg[&\mathrm{cos}(\pi((n\,\mathrm{cos}\theta_0+m\,\mathrm{sin}\theta_0)\,\tau-(n\,x_0+m\,y_0))))\notag\\
    \times\,&\mathrm{sin}(\pi((-n\,\mathrm{sin}\theta_0+m\,\mathrm{cos}\theta_0)\,h)\notag\\
    \times\,&(\pi(-n\,\mathrm{sin}\theta_0+m\,\mathrm{cos}\theta_0))^{-1}\Bigg]_{-\sqrt{1-\tau^2}}^{\sqrt{1-\tau^2}},
\end{align}
where we can discard the term containing $\mathrm{cos}([...]\,h)$ as it vanishes under the symmetric bounds. Evaluating the bounds, we reach:
\begin{align}
        \phi(\tau) \simeq \,&\mathrm{cos}(\pi((n\,\mathrm{cos}\theta_0+m\,\mathrm{sin}\theta_0)\,\tau-(n\,x_0+m\,y_0))))\notag\\
        \times\,&\mathrm{sin}(\pi((-n\,\mathrm{sin}\theta_0+m\,\mathrm{cos}\theta_0)\,h_1)(\pi(n\,\mathrm{sin}\theta_0-m\,\mathrm{cos}\theta_0))^{-1},
\end{align}
where $h_1 = \sqrt{1-\tau^2}$. Finally, we make the approximation that $h_1$ is slowly-varying compared to $\tau$ for most of the ingress/egress and thus its factor modulates $\phi$ by a factor that is on average $\frac{1}{\sqrt{2}}$, giving:
\begin{align}
    \phi(\tau) \simeq -\frac{1}{\sqrt{2}}&\mathrm{sinc}(\pi((n\,\mathrm{sin}\theta_0-m\,\mathrm{cos}\theta_0))\notag\\
    \times\,&\mathrm{cos}(\pi((n\,\mathrm{cos}\theta_0+m\,\mathrm{sin}\theta_0)\,\tau-(n\,x_0+m\,y_0)))).
\end{align}
We can see that the first part forms a constant prefactor, and the second a (co)sinusoidal variation in time, with phase given by the phase factors $x_0$ and $y_0$ (as expected, they do not affect the amplitude of the signal). As the prefactor is a $\rm sinc$, when its argument is near zero, $\phi$ will be order-1 for all $N$, however it will drop as $\sim 1/N$ for a poorly-aligned $(n,m)$ pair. Based on this, we can define $\theta: n=N\,\mathrm{cos}\theta,\,\, m=N\,\mathrm{sin}\theta$ (see Figure \ref{fig:eclipsemapping}B; this is why we picked our definition for $N$ earlier) and our phase factor $\gamma_p=-n\,x_0-m\,y_0$, and rewrite our formula as
\begin{equation}\label{eq:aligned}
    \phi(\tau)\simeq -\frac{1}{\sqrt{2}}\mathrm{sinc}(\pi\,N\, \mathrm{sin}(\theta-\theta_0))\,\mathrm{cos}(\pi(N\,\mathrm{cos}(\theta-\theta_0)\,\tau+\gamma_p))
\end{equation}
This form renders the notion of well- and poorly-aligned modes very clear (see Figure \ref{fig:eclipsemapping}C). When $\mathrm{sin}(\theta-\theta_0) < 1/N$, the mode is well-aligned and the differential light curve is order 1. When $\mathrm{sin}(\theta-\theta_0) > 1/N$, the amplitude of the differential phase curve drops rapidly with worsening alignment, and takes a typical value of $\sim1/N$. As there are $\sim N$ modes of order $N$, we can also see that there will always be a well-aligned mode, which will dominate the observed signal, especially for larger $N$. Note that the zeroes of the $\rm sinc$ do not necessarily have any physical correspondence, they merely arise from our choice of surface modes. The scaling and width of the maximum, however, are physical; these tell us about how bright well-aligned features will be relative to their poorly-aligned counterparts, and what angular range a feature must be in to be well-aligned.

Despite the notion of well-aligned/poorly-aligned modes being linked to the null space (in that well-aligned modes will dominate the non-null space if we choose to use the minimum $B(x,y)$ that can create a given light curve), it is important to point out that neither the well-aligned modes nor the poorly-aligned modes are entirely in nor out of the null space; the null space will be some linear combination of well-aligned and poorly-aligned modes, if the contributions from well-aligned modes to $B$ may be minor corrections.

While the argument from polynomial light curves only has two distinguishable modes per order (the inverting and non-inverting modes), the sinusoidal brightness distributions have four. This is because the even and odd polynomials are different orders, whereas their equivalents here - sine and cosine - are grouped into the same order. Hence the phase factor $\gamma_p$ links four separate modes; we could take them to be $\gamma_p = 0\,\, \text{or\,\,} \frac{\pi}{2}$, and either add $0 \,\, \text{or}\,\,\pi$ for the egress depending on whether the curve is inverting or not. 

\subsection{The full light curve}
Up until now, we have been dealing with the differential light curve as it is easier to calculate. We now have a form which we can integrate:
\begin{align}
    \Phi(\tau)&=\int_{-1}^{\tau} \phi(t_1)\,\mathrm{d}t_1 \notag\\
    &\simeq -\frac{1}{\sqrt{2}}\mathrm{sinc}(\pi\,N\,\mathrm{sin}(\theta-\theta_0))\notag\\
    &\quad\times\,\int_{-1}^{\tau}\mathrm{cos}(\pi(N\,\mathrm{cos}(\theta-\theta_0)\,t_1+\gamma_p)\,\mathrm{d}t_1.
\end{align}
Since we are only interested in the variations in $\Phi$ (we can assume we know the overall brightness of the planet from the eclipse depth), we can drop the lower bound of the integral, and, by taking the root-mean-square average, we get rid of phase information. Thus the scale of the variation in $\Phi$ is
\begin{equation}\label{eq:phiscale}
    \langle\Phi\rangle \simeq \frac{\mathrm{sinc}(\pi\,N\,\mathrm{sin}(\theta-\theta_0))}{2\,\pi\,N\,\mathrm{cos}(\theta-\theta_0)}.
\end{equation}
For well-aligned modes where the $\rm sinc$ is order-1, $\langle\Phi\rangle$ scales as $1/N$, while for poorly-aligned modes, it scales as $1/N^2$. One might notice the $\mathrm{cos}(\theta-\theta_0)$ in the denominator and suggest that perhaps some poorly-aligned modes will have significant contributions after all, but we should remember that $0 \leq \theta \leq \frac{\pi}{2}$ and $0 \leq \theta_0 \leq \frac{\pi}{2}$: only in rare cases when $\theta \simeq 0$ and $\theta_0 \simeq \frac{\pi}{2}$ (or vice-versa) will this factor become significant; for most impact parameters, the term will remain relatively small. We also note that the divergence here only arises due to the frequency vanishing, in which case the mode becomes degenerate with (much brighter) low-order modes, and our approximations of $h_1$ being slowly-varying break down. In short, we should not expect any poorly-aligned modes to become significant.

So while the brightness distributions are highly degenerate, with $\sim N^2$ coefficients constrained by only $\sim N$ equations, we find that only $\sim N$ modes tend to dominate the light-curve. Because of this, we expect that features produced by these well-aligned modes can be resolved fairly easily, as they yield large signals, while features produced by the poorly-aligned modes yield very weak signals and cannot be detected. We have however made approximations to reach these results, and the accuracy of our predictions will determine whether these approximations are reasonable.

\subsection{The final form}
We are now ready to estimate the signal-to-noise ratio expected for a given mode. The SNR is:
\begin{equation}
    \text{SNR} = \frac{F}{\sigma_F}
\end{equation}
where $\sigma_F$ is the error in $F$, and $F$ is the signal from the particular mode we are interested in, of order $N$. Assuming the measurements are independent and have the same error, $\sigma_F^2 = \sigma_1^2 N_F$ where $N_F$ is the number of measurements integrated together and $\sigma_1$ is the error per point. $F$ can also be estimated as $F = F_N\langle\Phi\rangle N_F$, where $F_{N} = B_N\,\pi R_p^2$; $B_N$ is the typical coefficient of modes at order $N$ in the expansion of $B(x,y)$, our brightness distribution. Hence:
\begin{equation}
    \text{SNR} = F_N\frac{\langle\Phi\rangle}{\sigma_1}\sqrt{N_F}.
\end{equation}
For regular measurements separated by a period $t_F$,
\begin{equation}
    N_F = \frac{2\,\Delta t}{T} \frac{T}{t_F}
\end{equation}
where $\Delta t$ is the duration of the ingress or egress, and the factor of $2$ comes from a full eclipse including both an ingress and egress. $T$ is the orbital period. Defining $N_\text{points} = T/t_F$, the number of measurements that would be made in a full orbit, and approximating $\Delta t$ as $\frac{R_p}{\pi a}\,T\, \mathrm{sec}\theta_0$,
\begin{equation}
    \text{SNR} = F_N\frac{\langle\Phi\rangle}{\sigma_1} \sqrt{\frac{2R_p}{\pi\, a \,\mathrm{cos}\theta_0}N_\text{points}}.
\end{equation}
We can use this approximation for $\Delta t$ as long as $1 - b >> R_p/R_s$, which we assumed earlier so we could take $\theta_0$ to be constant. We can define $\sigma = \sigma_1/\sqrt{N_\text{points}}$, the error in the flux from the planet as integrated over an entire orbit. Plugging in our value of $\langle\Phi\rangle$ from \ref{eq:phiscale} and considering only the contribution from the well-aligned modes (where $\theta \approx \theta_0)$,
\begin{equation}\label{eq:snr}
    \mathrm{SNR} = \frac{F_N}{\sigma} \frac{1}{\pi\,N} \sqrt{\frac{R_p}{2\pi\, a\, \mathrm{cos}\theta_0}}.
\end{equation}
We will assume the features to have a spatial power spectrum of $F_N \sim N^{-\gamma}$ where $\gamma$ is the gradient of the power spectrum. Our analytic formulation is agnostic to the value of $\gamma$, but to calculate estimates of the mapping resolution of real planets we assume $\gamma\approx2$ based on the feature scale in simulations of hot Jupiters \citep{cho2003changing}. By taking $\text{SNR} \simeq 1$ at the limit of detectability, $F_N \simeq 2F_{0}\, N^{-\gamma}$, and then inverting this equation, we can calculate $N_{\text{max}}$, the highest order detectable:
\begin{equation} \label{eq:nmax}
    N_{\text{max}} = \left[\frac{F_0}{\sigma} \frac{1}{\pi} \sqrt{\frac{2R_p}{\pi\, a\, \mathrm{cos}\theta_0}}\,\,\right]^\frac{1}{\gamma+1},
\end{equation}
where $F_0$ is the phase-averaged flux from the planet and $\sigma$ is the error on the phase-averaged flux as integrated over an entire orbit. The factor of $2$ in $F_N$ arises from the day-side flux being typically substantially larger than the night side, and thus the phase-averaged flux is around half the total day-side flux. Note that this falloff is better than inverting the phase curve, where the exponent is $\frac{1}{\gamma+2}$ (see \citet{cowan2008inverting}). This is because the brightness contributed by a given patch of planetary surface drops discontinuously to zero as it passes behind the star, whereas it falls off linearly as the planet rotates. If we follow \citet{cowan2013light} and look at the effect of convolving our brightness distributions with these (multiplying the power spectra), the step (for eclipse maps) scales as the inverse of wavenumber, while the linear increase (for phase curves) scales as the inverse square of the wavenumber. While the exponent is better for eclipse mapping, phase curve inversion benefits from a longer integration time (the whole orbit compared to just the eclipse ingress/egress), so may outperform eclipse mapping when $F_0/\sigma$ is lower. The threshold is around
\begin{equation}
    \frac{F_0}{\sigma} \approx \left(\frac{\Delta t}{T}\right)^{-\frac{\gamma+2}{2}},
\end{equation}
where $\Delta t$ is the duration of the ingress/egress and $T$ the orbital period. This is only accurate to within an order-unity factor, since the modes used here and in phase curve inversion are not directly comparable. 

Despite each individual mode's strong $\theta_0$ dependence, $N_{\text{max}}$ only depends on $\theta_0$ insofar as it affects the duration of the ingress/egress; as long as there is a well-aligned mode, it does not matter which mode that is.

\subsection{Resolution at a given order}\label{sec:resolution}
Consider a brightness distribution dominated by a single sharp peak which integrates to $1$. The corresponding differential light curve must also share this peak. Without loss of generality, we will place it at $(0,0)$, the centre of the planetary disk, which also places it at $\tau=0$ in $\phi(\tau)$. Fitting this differential light curve using the well-aligned modes ($\theta=\theta_0$) up to order $N$, we have:
\begin{equation}
    \phi \simeq \delta(\tau) \simeq \sum_{k=0}^N \, A_k \,\mathrm{cos}(\pi \,k \,\tau),
\end{equation}
where we are approximating the sharp peak as a Dirac delta function, similar to Fig. 2 of \citet{cowan2017mapping}. The coefficients $A_k$ are the Fourier coefficients of the delta function, $A_0 = \frac{1}{2}, A_{k \neq 0} = 1$. Let us estimate the size of the resulting brightness peak using the second derivatives of $B$ with respect to $x$ and $y$ around the origin:
\begin{gather}
    B(x,y) = \sum_{k=0}^N \, A_k \, \mathrm{cos}\left(\pi\,k\left(x\mathrm{cos}\theta_0 + y\mathrm{sin}\theta_0\right)\right)\\
    \frac{\partial^2}{\partial x^2} B 
\biggr\rvert_{x,y = 0} = \sum_{k=0}^N\,\left(\pi \, k\, \mathrm{cos}\theta_0\right)^2,
\end{gather}
and similar for $y$ with $\mathrm{sin}\theta_0$ instead of $\mathrm{cos}\theta_0$. This sum evaluates to:
\begin{equation}
    \frac{\partial^2}{\partial x^2} B \biggr\rvert_{x,y = 0} = \frac{1}{6}\,N\,(N+1)\,(2N+1)\,\left(\pi\, \mathrm{cos}\theta_0\right)^2.
\end{equation}
The height of the peak is $N+\frac{1}{2}$, and approximating the peak using a Gaussian and matching the second derivative at the origin and the height of the peak, we find:
\begin{equation}\label{eq:resolution}
    \begin{split}
        B(x,0) \simeq \left(N+\frac{1}{2}\right) \, \mathrm{exp}\left(-\frac{x^2}{2 x_r^2}\right)\\
        \text{where}\,\, x_r = \frac{1}{\sqrt{2}\pi\,\mathrm{cos}\theta_0}\sqrt{\frac{6}{N(N+1)}}.
    \end{split}
\end{equation}
Similarly, we can define $y_r$ as follows:
\begin{equation}\label{eq:yresolution}
    y_r = \frac{1}{\sqrt{2}\pi\,\mathrm{sin}\theta_0}\sqrt{\frac{6}{N(N+1)}}.
\end{equation}
For a Gaussian of standard deviation $x_{r}$, the full width at half maximum is $X_{r} = 2\sqrt{2 \ln 2}\ x_{r}$. Then the resolved spot covers an angular range of $r_{x} = 2\,\arcsin(X_{r}/2)$. Note that this estimate for $r_x$ is only accurate for a spot in the centre of the planetary disk (at the substellar point), and the angular resolution worsens for features towards the edge of the disk as the planetary surface curves away from the line of sight. In the limit of large $N$ ($N > \sim 4$), $x_r$ scales as $(N\,\mathrm{cos}\theta_0)^{-1}$ with a prefactor of $\frac{\sqrt{3}}{\pi}$. For the ideal resolution, $N=N_{\text{max}}$, with $N_{\text{max}}$ from Equation \ref{eq:nmax}. 

Thus, based on the eclipse light curve alone, we cannot resolve features smaller than $\sim r_x$ in longitude and $\sim r_y$ in latitude. Since $r_x$ diverges as $\theta_0 \to \frac{\pi}{2}$ and $r_y$ diverges as $\theta_0 \to 0$, we have a tradeoff between longitude and latitude resolution; unless we have an excellent SNR, we cannot expect to have good resolution in both, and picking a target with good latitude resolution will mean compromising on longitude resolution.

Unfortunately, while Equations \ref{eq:nmax} and \ref{eq:yresolution} tell us that a high stellar edge angle $\theta_0$ grants us both a better SNR via longer ingress/egress durations and a better resolution along the latitude direction (which cannot be constrained via phase curves), planets are roughly evenly spaced in impact parameter $b = \mathrm{sin}\theta_0$ and thus planets with good latitude resolution will be few and far between compared to those with good longitude resolution; see Figure \ref{fig:resolution}, where the top $20\%$ of impact parameters correspond to the top $40\%$ of stellar edge angles. This makes the few bright planets with high impact parameters ($b \gtrsim 0.7$) very valuable for fitting maps with good two-dimensional resolution.

\subsection{Eclipse Mapping Metric}
Based on our theory, we propose a simple metric which approximates (and is monotonic in) the resolution attainable using eclipse mapping. We combine Equation \ref{eq:nmax} with $\gamma=2$, Equation \ref{eq:resolution}, and subsequent calculations to give:
\begin{align}\label{eq:EMM}
    \text{EMM}_x  &= 180^\circ\times\left[\frac{2\sqrt{6 \ln 2}}{\pi^2\sqrt{1-b^2}}\left(\left(\frac{F_0}{\sigma} \frac{1}{\pi} \sqrt{\frac{2R_p}{\pi\, a\, \sqrt{1-b^2}}}\,\,\right)^{\frac{1}{3}}+\frac{1}{2}\right)^{-1}\right]\notag\\
    \text{EMM}_y  &= 180^\circ\times\left[\,\,\,\frac{2\sqrt{6 \ln 2}}{\pi^2\,\,\,b\,\,}\,\,\,\left(\left(\frac{F_0}{\sigma} \frac{1}{\pi} \sqrt{\frac{2R_p}{\pi\, a\, \sqrt{1-b^2}}}\,\,\right)^{\frac{1}{3}}+\frac{1}{2}\right)^{-1}\right]\notag\\
    \text{EMM}_{\,\,} &= \,\,\sqrt{\text{EMM}_x^2 + \text{EMM}_y^2}.
\end{align}
$\text{EMM}$ is a combined resolution which estimates the diagonal extent of a resolution element. Other parametrisations for $\text{EMM}(\text{EMM}_x,\text{EMM}_y)$ are possible and may be useful for certain observations. We remind the reader that in these expressions, $F_0$ is the flux from the planet averaged over one orbital phase, and $\sigma$ is the corresponding error, i.e. $\sigma = \sigma_1/\sqrt{N_\text{points}}$ where $\sigma_1$ is the error for a single data point and $N_\text{points}$ is the number of points in a complete orbit. As above, $b$ is the impact parameter, $R_{p}$ is the planetary radius, and $a$ is the semi-major axis.

\begin{figure}
    \centering
    \includegraphics[width=\columnwidth]{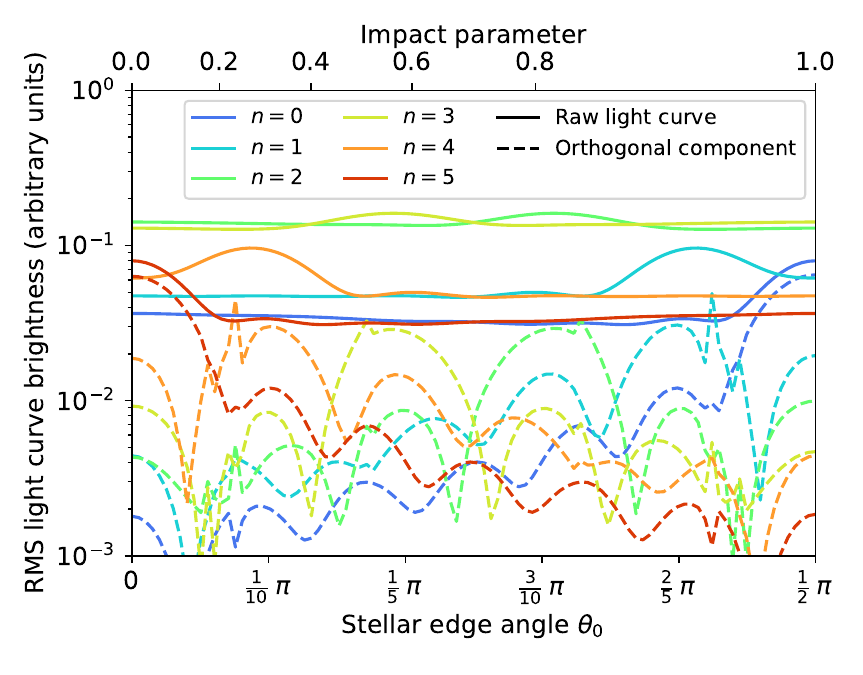}
    \caption{The RMS brightness for the order $N=5$ light-curves and their orthogonal components as a function of stellar edge angle. The orthogonal components are calculated by orthogonalising each curve with respect to the lightcurves with $N<5$, so they are not orthogonal to each other. While the non-orthogonalised light curves have a high baseline intensity, this is contributed by low-frequency components which are degenerate with the lower-order modes. The peaks are from the high-frequency component; since this component cannot be found in the lower-order modes, it also remains in the orthogonalised curves. We also note that the peaks in the orthogonal components are several times higher than the typical background, which matches the theory which suggests that the well-aligned modes should be $\sim N$ times brighter than the background.}
    \label{fig:orthogs}
\end{figure}

\section{Numerical analysis}\label{sec:methods}
We implement a system to fit a brightness distribution from a secondary eclipse light curve. Our aim is to produce a numerical fit to a light curve that matches the analytic process described above, in order to test our analytic predictions of mapping resolution.

We first generate the light curves corresponding to the well-aligned modes up to an order $N$ (defined here as $n+m=N$, for easier comparisons between modes), as well as the poorly-aligned modes up to the same order. There are only two modes whenever one of $n, m$ is zero (in this situation, the phase factor controls the amplitude and the phase-shifted curves are zero), and so we borrow the phase-shifted modes from the next $n,m$ up by adding 1 to whichever of $n,m$ is zero, and these modes are then removed from the set of poorly-aligned modes. These should still be on the central peak of the $\rm sinc$, if lower down, so they should still be reasonably bright. We orthogonalise all the modes using the Gram-Schmidt procedure, running all the well-aligned modes first, in order of increasing order $N$. The well-aligned modes do not overlap significantly so their orthogonalised versions are mostly unchanged; the orthogonalised poorly-aligned modes vanish due to the degeneracy, so we are left with a basis for the light curve, and a `null space' of brightness distributions which contribute zero to the light curve. 

\begin{figure*}
    \centering
    \includegraphics[width=\textwidth]{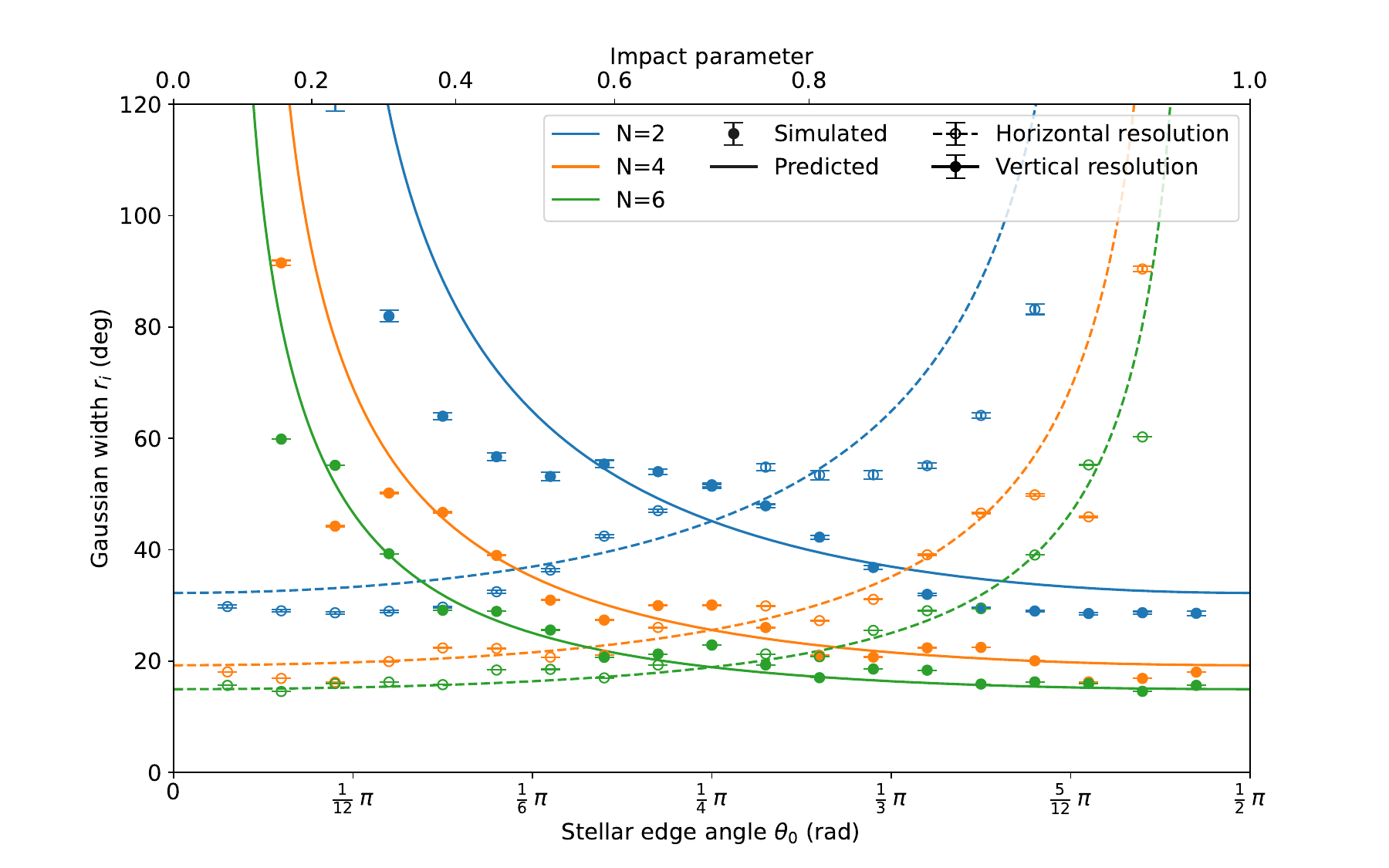}
    \caption{Horizontal and vertical resolutions as a function of stellar edge angle $\theta_0$, for three different fitting orders $N$. We estimate the Gaussian widths using the second derivative of the brightness at the origin. The error bars are calculated using the posterior distribution from the MCMC fit. The input brightness distribution is a radially symmetric Gaussian with width $0.1\,R_p$. Because the $N=6$ simulations had resolutions comparable to the size of the Gaussian input, the predicted curves are calculated as $\sqrt{i_{r}(\theta_0)^2 + 0.1^2}$ where $i = x,y$ (see Equation \ref{eq:resolution} for the definitions of $x_r$ and $y_r$). We note that the simulated data oscillate around the predicted values with frequency proportional to the order $N$; we do not as yet have an explanation for this. We also note the skew between impact parameter $b$ and stellar edge angle $\theta_0$; almost 40\% of the range of stellar edge angles corresponds to impact parameters greater than 0.8.}
    \label{fig:resolution}
\end{figure*}

We use a Markov Chain Monte-Carlo (MCMC) to fit the coefficients of the basis light curves to the input curve, and generate a posterior distribution of coefficients. To fit the poorly-aligned modes, we borrow from the `slice mapping' method of \citet{majeau2012two} and assume that the true distribution is the smoothest possible distribution for that particular combination of well-aligned modes; we use a least squares solver to maximise the smoothness of the map. While smoothness maximisation is not generally realistic, it does have the result of removing features that are not constrained by the light curve (thus probably spurious) while retaining those which are well-constrained. 

When fitting the light-curves for Figure \ref{fig:order}, as we are not interested in the distributions of parameters that correspond to any given light-curve (or the resulting maps), we calculated the coefficients for the modes by taking the scalar product of the input light-curve with the orthogonalised well-aligned mode light curves.

It is worth noting two things. First, that while we use the well-aligned modes for fitting the light curve, the final map's light curve will draw from some mixture of well- and poorly-aligned modes, and the same map should be reached via any method which breaks the degeneracy by optimising smoothness. Second, the map will be incorrect, as the real map will not in general maximise smoothness. Smoothness maximisation does however provide us with a way to break the degeneracy while yielding reasonable values; it mostly preserves large features, while removing usually spurious smaller oscillations.

We use RK4 integration assuming $R_p << R_s$ (straight line stellar edge) to generate the differential light-curve, which we then integrate with RK4 again. We do not include the planet's rotation from ingress to egress.  To impose smoothness, we calculate the fitted brightness distribution on a $50\times50$ grid, and define smoothness per grid point as the sum of the squared differences between the grid point and its four neighbours. The total smoothness is then the sum over the smoothness of all the grid points in the planetary disk. Note that the smoothness score defined here is minimised when the solution is most smooth.

\section{Results}\label{sec:results}
Our analytic theory makes three main predictions. Firstly, the space of brightness distributions is highly degenerate with respect to light curves. Secondly, that the SNR at a given order $N$ goes as $N^{-1}$ multiplied by the intrinsic brightness of the modes of that order. Finally, the achievable resolutions along the horizontal and vertical axes are related to the angle of the stellar edge with respect to the planet's motion as calculated in Equation \ref{eq:resolution}. None of these are particularly unexpected; degeneracy in eclipse mapping is discussed even prior to the discovery of the first exoplanets \citep{horne1985images} and in exoplanetary contexts such as \citet{rauscher2007toward, challener2023eclipse}; the scaling can be derived fairly easily from \citet{cowan2013light}; and that a low impact parameter implies little to no latitude resolution can be observed from symmetry considerations. Nonetheless, we can find neither the scaling nor the precise variation of zonal and meridional resolution with impact parameter explicitly stated in the literature. While \citet{adams2022sensitivity} mention the angle of the stellar edge as a determining factor in what features are resolvable, they do not elaborate on this.

\begin{figure}
    \centering
    \includegraphics[width=\columnwidth]{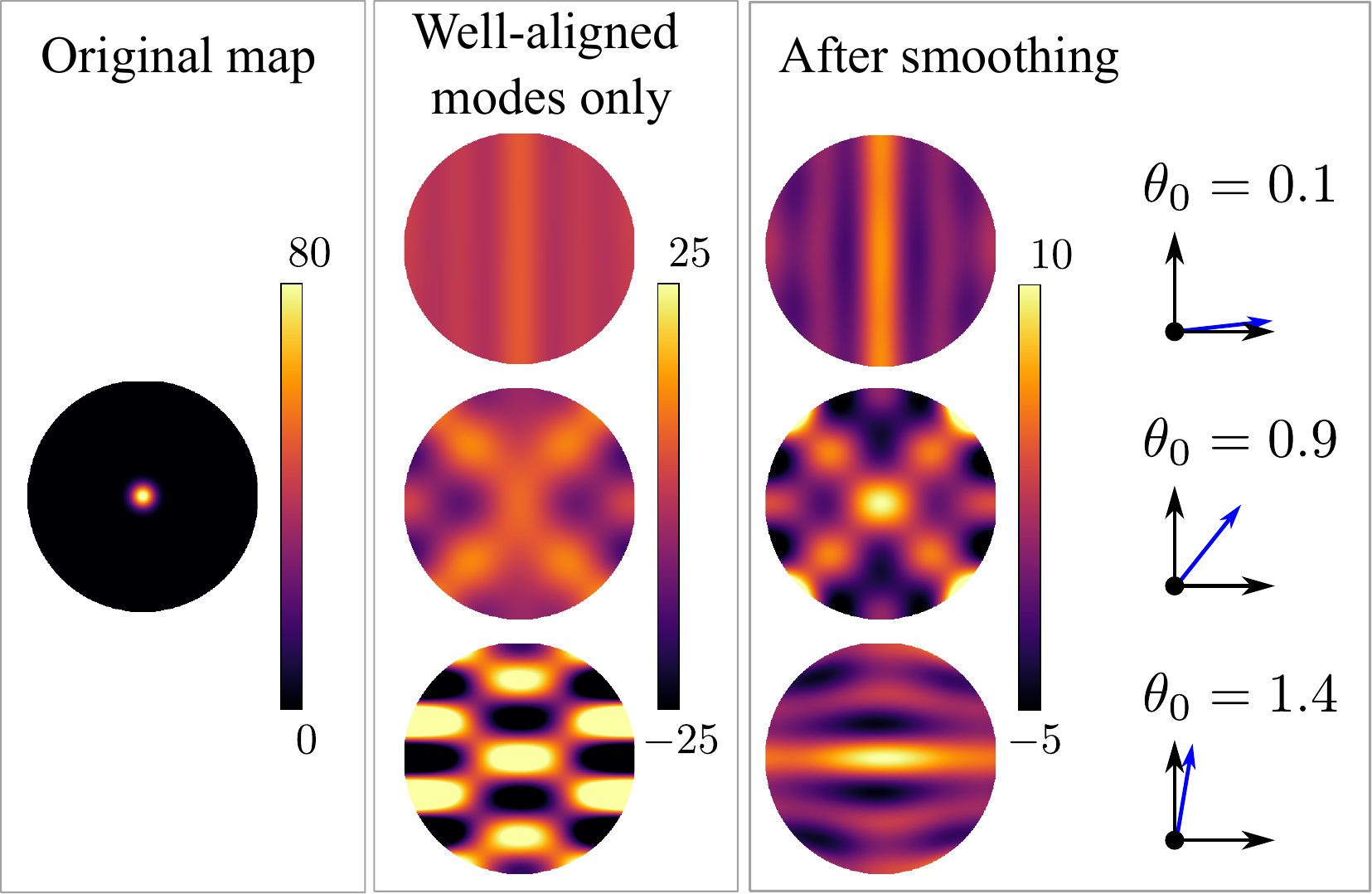}
    \caption{The method described in \ref{sec:methods}, as applied to the Gaussian map used for Figure \ref{fig:resolution}. We have intentionally picked a very small hot-spot in the input map so as to approximate a delta-function, so the results can be more easily compared with our theory in Section \ref{sec:resolution}. When the well-aligned maps are already fairly smooth (top), there is little change during the smoothing procedure. However, if the well-aligned map is less smooth, or if one of the modes selected as well-aligned is on the edge of being poorly-aligned (bottom), the smoothing procedure has a larger effect and generally does well in retrieving a reasonable result. We used a fitting order $N=4$ for these tests.}
    \label{fig:gaussian_maps}
\end{figure}
\begin{figure}
    \centering
    \includegraphics[width=0.9\columnwidth]{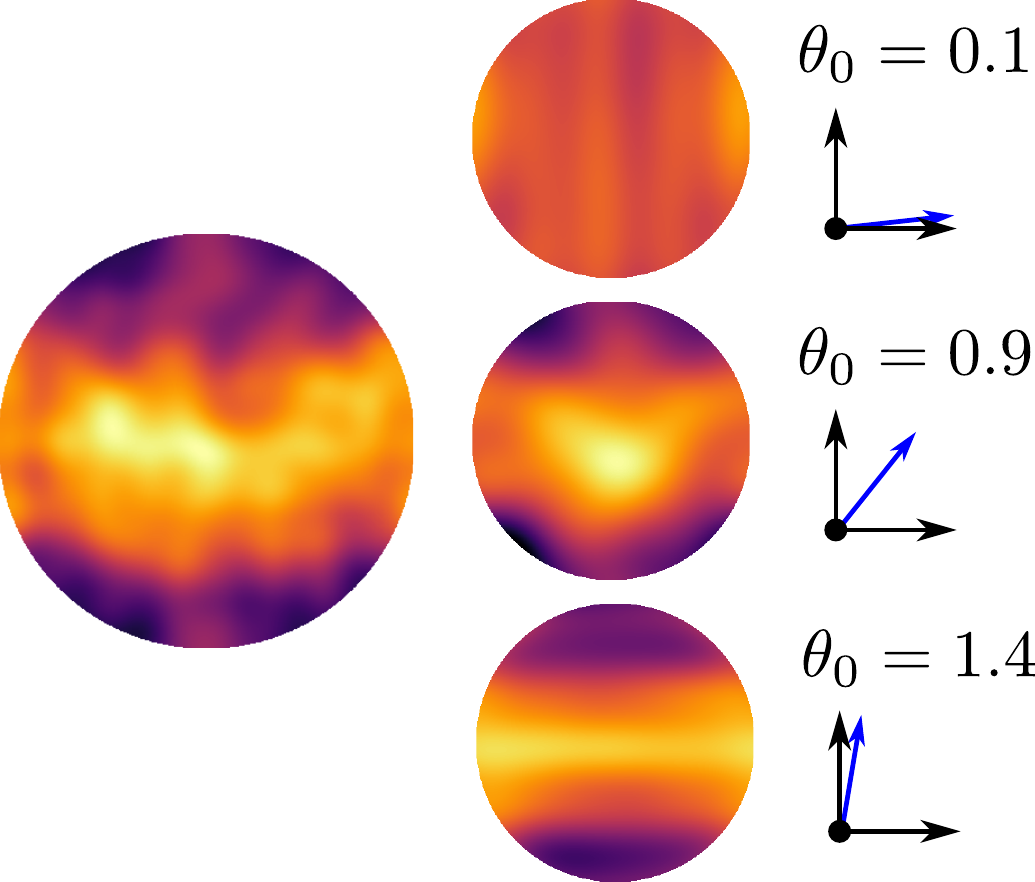}
    \caption{A randomly generated map (left) and the retrieved maps for fitting order $N=4$ at three different values of $\theta_0$ (right). At $\theta_0=0.1$, the horizontal band is very poorly-aligned and so is not detected; the smoothing process then prefers a smooth map with few features. At $\theta_0=0.9$, the horizontal band can be detected, alongside variations along $x$. Finally, at $\theta_0=1.4$, the horizontal band is well-aligned but variations in the $x$ direction cannot be detected and are smeared out. Around $\theta=\frac{\pi}{4}$, the smoothing cannot distinguish between horizontal and vertical features as long as they are symmetric across both axes, so turns the horizontal band into a plus shape; in real situations, these features could be distinguished from one another using the phase curve, so we picked a stellar edge angle reasonably close by ($\theta_0=0.9$) which does not suffer from this problem.}
    \label{fig:maps}
\end{figure}

When orthogonalising the light-curves, we find that for fitting order $N=4$, the average RMS for the orthogonalised well-aligned modes is usually a factor of a hundred or more larger than the RMS of the brightest orthogonalised poorly-aligned mode. The reason the null space curves are not reduced all the way to zero is that each light-curve is composed of many higher frequencies (especially at the beginning and end of the ingress and egress curves where there is a discontinuity in the second derivative), and these components, while small, are different for each light-curve. Figure \ref{fig:orthogs} shows the $N=5$ light curves and their components orthogonal to all the $N < 5$ modes. While the low-frequency term dropped in Equation \ref{eq:twofreq} contributes a roughly constant baseline to each mode which dominates the signal (solid lines), the high-frequency component (not found in the $N<5$ modes) dominates the orthogonal components (dashed lines). We see the peaks predicted by Equation \ref{eq:aligned}, and that there is always a well-aligned mode. 

To demonstrate that the system described in \ref{sec:methods} works reliably, we ran several tests giving the system known maps and seeing how well it could recreate them from the light curve. As expected, the method is able to fairly reliably retrieve features which are reasonably well-aligned with the stellar edge during the eclipse. 

\begin{figure}
    \centering
    \includegraphics[width=0.9\columnwidth]{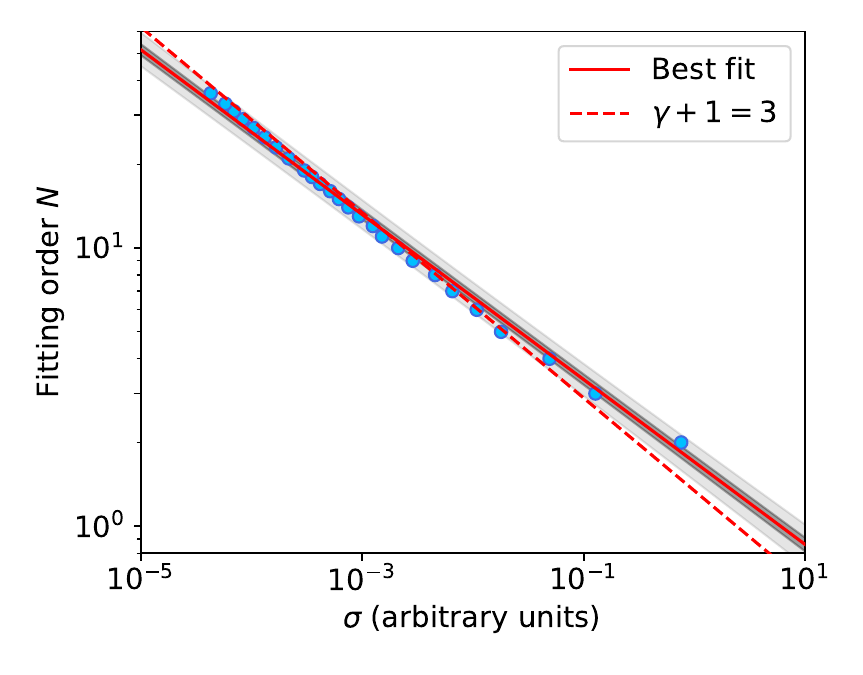}
    \caption{Best fitting order $N$ as a function of $\sigma$. We find $\sigma(N)$ numerically by fitting a known light-curve with noise added; we calculate the value of $\chi^2$ for the fit as compared to the noiseless curve. $\sigma(N)$ is then the value of $\sigma$ for which $\chi^2 \sigma$ is minimised. We chose this score as the mean squared error $\chi^2 \sigma^2$ is sensitive to overfitting but not underfitting, while $\chi^2$ is sensitive to underfitting but not overfitting. Their product then is sensitive to both. From Equation \ref{eq:nmax}, we expect $\sigma(N) \sim N^{\gamma+1}$ if the brightness of the modes of order $N$ follows $N^{-\gamma}$. In this case we used $\gamma=2$ and find our best fit from $\sigma(N)$ at $\gamma+1 = 3.37 \pm 0.05$, reasonably close to the predicted value. The best fit gradient also drops closer to the prediction if we exclude the lower-order points. For phase curve inversion, the power law would have an exponent of $\gamma+2$ and thus the same number of data points would constrain fewer surface features. The dark and light grey bands around the best fit line represent the $1\sigma$ and $3\sigma$ bounds respectively. We used $\theta_0=\frac{\pi}{4}$ for these runs.}
    \label{fig:order}
\end{figure}

In Figure \ref{fig:gaussian_maps}, we test our method's ability to retrieve what is effectively a delta-function map (we use $N=4$, which is unable to resolve the small hot-spot), and hence is directly comparable to our theory in Section \ref{sec:resolution}. As expected, the resulting maps have large brightness peaks at the correct position, but whose sizes along $x$ and $y$ depend strongly on the stellar edge angle $\theta_0$, being stretched perpendicular to the stellar edge. 

Figure \ref{fig:maps} shows the input map and maps retrieved from light-curves generated with three different stellar edge angles (marked by the green lines). When the stellar edge is almost vertical, the light curve does not distinguish between a narrow equatorial band and a uniformly warm planet, and so the smoothing routine prefers the less-detailed map with no equatorial band. When the stellar edge is more favourable for resolving horizontal features, the equatorial band is correctly retrieved, as is the presence of a hot-spot, though the exact position is slightly offset. For an almost horizontal stellar edge, the horizontal band is easily retrieved but the position of the hot-spot in longitude is lost. Note however that including phase curve data will give some longitude information (though as found in Equation \ref{eq:nmax} compared with \citet{cowan2008inverting}, the phase curve is much less sensitive to small features); combining these is beyond the scope of this paper.

Figure \ref{fig:resolution} shows how we applied the method to fit Gaussian maps using different SNRs and correspondingly different values of fitting order $N$. We find that the widths of the fitted Gaussians along the horizontal and vertical axes trace the values predicted by Equation \ref{eq:resolution} very closely, with RMS errors of $5,3,2^\circ$ over the $|\theta-\theta_0| < 60^\circ$ range for the $N=2, 4, 6$ tests respectively, each corresponding to an average error of 10\% compared to the predicted values. 

We also ran several tests fitting randomly-generated maps where the coefficient of each mode is calculated from a standard normally-distributed variable multiplied by $N^{-\gamma}$, where we used $\gamma=2$. From the theory (Equation \ref{eq:snr}), we expect that the value of $\sigma$ required to be able to justify adding another order should go as $N^{-\gamma-1}$; in this case $N^{-3}$. Using \verb|scipy.optimize|'s curve-fitting routine, we find a best fit value of $\gamma+1=3.37 \pm 0.05$ (see Figure \ref{fig:order}), reasonably close to the predicted value. Most of the deviation comes from low orders which are more significantly discrepant from $\gamma+1=3$, and fitting only the higher orders ($N \geq 10$) yields $\gamma+1=3.02\pm0.01$. While we use the well-aligned modes to perform our fitting and one might suggest that this would alter the results (since Equation \ref{eq:nmax} only considers the well-aligned modes), all bases with identical span would yield the same light-curves and thus the same errors and power law.

\section{Real Planets}\label{sec:real_planets}

\begin{figure}
    \centering
    \includegraphics[width=\columnwidth]{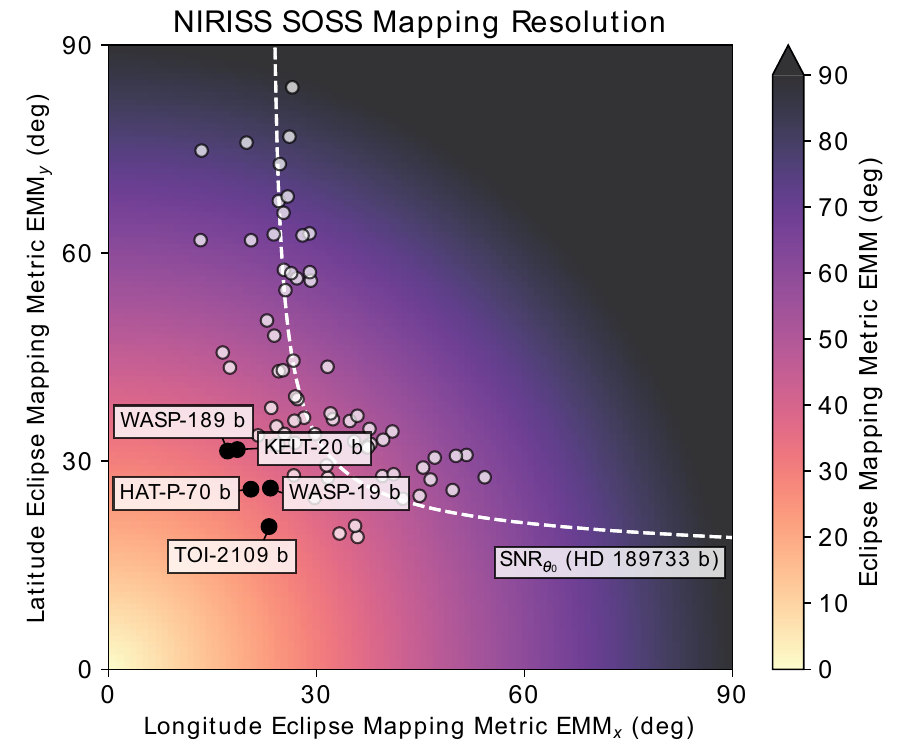}
    \caption{The best 100 targets for eclipse mapping with NIRISS SOSS, selected as described in Section \ref{sec:real_planets}.  The five best targets by single-eclipse overall resolution EMM according to Equation \ref{eq:EMM} are labelled.}
    \label{fig:niriss_res}
\end{figure}

\begin{figure}
    \centering
    \includegraphics[width=\columnwidth]{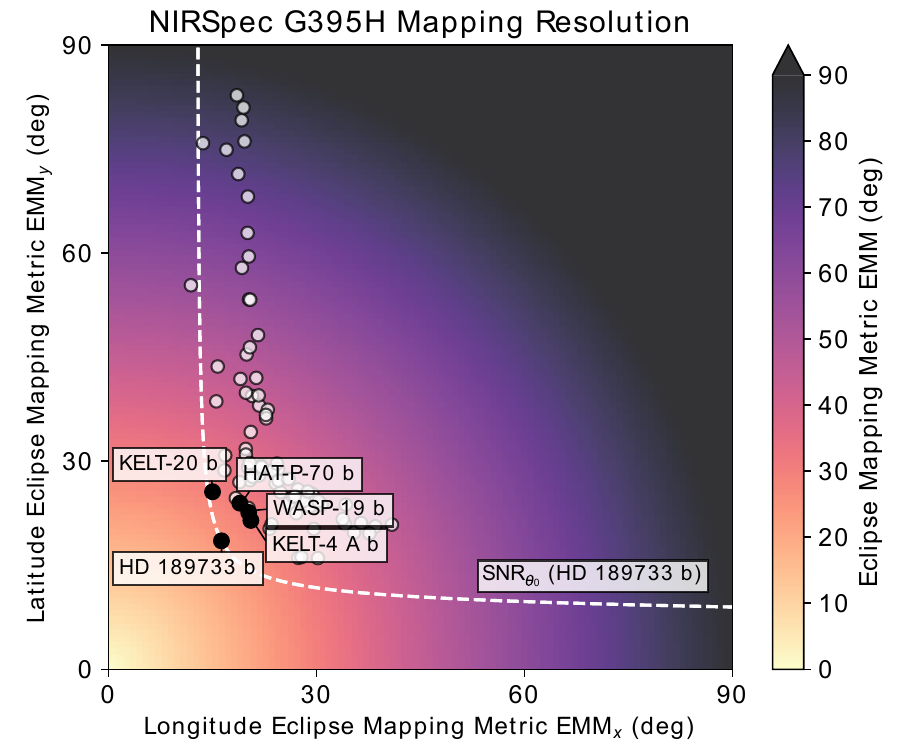}
    \caption{The best 100 targets for eclipse mapping with NIRSpec G395H, selected as described in Section \ref{sec:real_planets}.  The five best targets by single-eclipse overall resolution EMM according to Equation \ref{eq:EMM} are labelled.}
    \label{fig:nirspec_res}
\end{figure}

\begin{figure}
    \centering
    \includegraphics[width=\columnwidth]{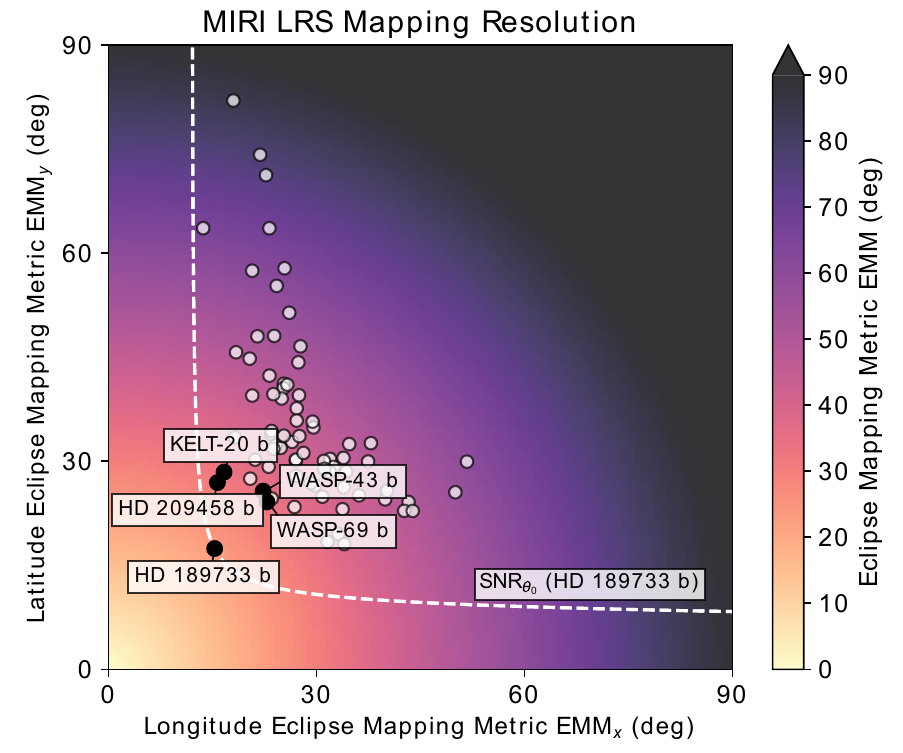}
    \caption{The best 100 targets for eclipse mapping with MIRI LRS, selected as described in Section \ref{sec:real_planets}. The five best targets by single-eclipse overall resolution EMM according to Equation \ref{eq:EMM} are labelled.}
    \label{fig:miri_res}
\end{figure}

Our analytic result in Equation \ref{eq:resolution} predicts the angular size of features resolvable by eclipse mapping of a planet given the parameters of its system and the signal to noise ratio of the observations. Given this, we can derive the expected resolution of an eclipse map of any planet with any instrument. We apply this to all known exoplanets for three instruments: MIRI LRS, NIRSpec G395H, and NIRISS SOSS. We do not include grazing orbits as they are not described by our theory, but some of these may be good targets for partial mapping.

We select the top 100 targets for each instrument using the more approximate and non-absolute eclipse mapping metric $\sim\ \frac{F_{p}}{F_{S}}  \sqrt{\frac{\Delta t}{\Delta t_{0}} 10^{0.4(K-K_{0})}}$ \citep{mullally2019exo}, where $\Delta t$ is the duration of eclipse ingress and egress and $K$ is the K-band magnitude of the observed star ($\Delta t_{0}$ and $K_{0}$ are the reference values for the planet HD 209458 b). We calculate the average flux ratio in the relevant wavelength band, and calculate the cadence and the error per point in a phase curve using Pandexo \citep{batalha2017pandexo}.

We then calculate the eclipse mapping metrics in Equation \ref{eq:EMM} for the top 100 targets selected in this way for each instrument. We estimate $F_{0}$ as the average flux ratio in the relevant wavelength band, where the planetary flux is calculated from its equilibrium temperature. Figures \ref{fig:niriss_res}, \ref{fig:nirspec_res}, and \ref{fig:miri_res} show the resulting longitudinal, latitudinal, and overall resolutions. The targets with the five best single-eclipse overall resolution estimates (EMM) are labelled by name. 

The dashed line in each figure shows the effect of varying the stellar edge angle on the expected mapping resolution for a planet with the SNR of HD 189733 b, where we include the duration of its eclipse in the SNR, defining it as $\mathrm{SNR}_{\theta_{0}}$. Figures \ref{fig:niriss_res}, \ref{fig:nirspec_res}, and \ref{fig:miri_res} show a skew towards better longitudinal resolution EMM$_{x}$ and worse latitudinal resolution EMM$_{y}$, with planets clustering in the top left of the plot. This is because if planets are uniformly distributed in impact parameter $b$, their stellar edge angle $\theta_0 = \mathrm{arcsin}(b)$ will be distributed non-uniformly in the range 0 to 90$^{\circ}$, with more small values than large values, resulting in EMM$_y$ being typically larger than EMM$_x$. 

Tables \ref{tab:niriss_targets}, \ref{tab:nirspec_targets}, and \ref{tab:miri_targets} list the top 15 targets for each instrument, ordered by the best overall resolution EMM. As would be expected, the shorter-wavelength NIRISS SOSS is better suited to mapping higher-temperature planets, while the longer-wavelength MIRI LRS is better suited to mapping lower-temperature planets. In general, an impact parameter around $0.5$ to $0.8$ (giving an intermediate stellar edge angle) is key for a good eclipse mapping signal. We have not included the effect of information from the out-of-eclipse phase curve when deriving an eclipse map. This provides more longitudinal information (albeit with less spatial resolution, as discussed above) than the eclipse only. Including phase curve information will therefore make planets with high $b$ where EMM$_{y}$ < EMM$_{x}$ better targets for eclipse mapping.

Table \ref{tab:niriss_targets} also lists the expected resolution of WASP-18 b as \citet{coulombe2023broadband} derived an eclipse map of WASP-18 b from an observed secondary eclipse with NIRISS SOSS, finding good constraints on its horizontal (longitudinal) structure but poor constraints on its vertical (latitudinal) structure. This is consistent with the analytic prediction of Table \ref{tab:niriss_targets} of EMM$_{x}=19.8^\circ$ and EMM$_{y}=52.6^\circ$ for this planet and instrument.



\begin{table*}
    \centering
    \begin{tabular}{rl|cccccccc}
\multicolumn{2}{c}{\textbf{Planet}}  &  \textbf{Equilibrium}  &  \textbf{Radius}  &  \textbf{Period} & \textbf{Impact}  & \textbf{K}  &  \textbf{EMM}$_{x}$  &  \textbf{EMM}$_{y}$ &  \textbf{EMM} \\
\multicolumn{2}{c}{\textbf{Name}}  &  \textbf{Temperature (K)}  &  ($R_{J}$)  &  \textbf{(days)} & \textbf{Parameter}& \textbf{(mag)} & \textbf{(deg)} &  \textbf{(deg)}  &  \textbf{(deg)} \\
\hline
&  TOI-2109 b  &  3072  &  1.35  &  0.67  &  0.75  &  9.07  &  23.1  &  20.6  &  31.0 \\
 &  HAT-P-70 b  &  2552  &  1.87  &  2.74  &  0.62  &  8.96  &  20.5  &  26.0  &  33.1 \\
 &  WASP-19 b  &  2117  &  1.42  &  0.79  &  0.67  &  10.48  &  23.3  &  26.1  &  35.0 \\
* &  WASP-189 b  &  2636  &  1.62  &  2.72  &  0.48  &  6.06  &  17.2  &  31.5  &  35.9 \\
* &  KELT-20 b  &  2255  &  1.74  &  3.47  &  0.51  &  7.42  &  18.6  &  31.7  &  36.7 \\
 &  KELT-14 b  &  1961  &  1.74  &  1.71  &  0.86  &  9.42  &  33.3  &  19.6  &  38.6 \\
 &  WASP-167 b  &  2363  &  1.58  &  2.02  &  0.77  &  9.76  &  29.7  &  24.7  &  38.6 \\
 &  KELT-4 A b  &  1821  &  1.7  &  2.99  &  0.69  &  8.69  &  26.7  &  28.0  &  38.7 \\
 &  WASP-178 b  &  2469  &  1.81  &  3.34  &  0.54  &  9.7  &  21.6  &  33.7  &  40.0 \\
 &  KELT-7 b  &  2042  &  1.6  &  2.73  &  0.6  &  7.54  &  24.1  &  32.4  &  40.3 \\
 &  WASP-87 b  &  2313  &  1.38  &  1.68  &  0.6  &  9.55  &  24.5  &  32.4  &  40.6 \\
 &  TOI-1431 b  &  2395  &  1.49  &  2.65  &  0.88  &  7.44  &  35.9  &  19.1  &  40.7 \\
 &  WASP-74 b  &  1916  &  1.36  &  2.14  &  0.86  &  8.22  &  35.6  &  20.7  &  41.1 \\
 &  KELT-8 b  &  1676  &  1.62  &  3.24  &  0.75  &  9.18  &  31.6  &  27.6  &  42.0 \\
 &  WASP-100 b  &  2195  &  1.33  &  2.85  &  0.64  &  9.67  &  26.9  &  32.6  &  42.3 \\
 ** &  WASP-18 b  &  2504  &  1.24  &  0.94  &  0.37  &  8.13  &  17.5  &  43.5  &  46.9 \\

    \end{tabular}
    \caption{The best 15 targets for eclipse mapping with NIRISS SOSS, ordered by their single-eclipse overall resolution EMM according to Equation \ref{eq:EMM}. The targets marked with single asterisks are expected to partially saturate the detector but we leave them here for reference. Additionally, the predicted resolution of WASP-18 b (marked with a double asterisk) is listed for comparison with the real eclipse map derived by \citet{coulombe2023broadband} with NIRISS SOSS.}
    \label{tab:niriss_targets}
\end{table*}

\begin{table*}
    \centering
    \begin{tabular}{rl|cccccccc}
\multicolumn{2}{c}{\textbf{Planet}}  &  \textbf{Equilibrium}  &  \textbf{Radius}  &  \textbf{Period} & \textbf{Impact}  & \textbf{K}  &  \textbf{EMM}$_{x}$  &  \textbf{EMM}$_{y}$ &  \textbf{EMM} \\
\multicolumn{2}{c}{\textbf{Name}}  &  \textbf{Temperature (K)}  &  ($R_{J}$)  &  \textbf{(days)} & \textbf{Parameter}& \textbf{(mag)} & \textbf{(deg)} &  \textbf{(deg)}  &  \textbf{(deg)} \\
\hline
*&  HD 189733 b  &  1202  &  1.13  &  2.22  &  0.66  &  5.54  &  16.3  &  18.5  &  24.7 \\
 &  KELT-20 b  &  2255  &  1.74  &  3.47  &  0.51  &  7.42  &  15.0  &  25.6  &  29.7 \\
 &  KELT-4 A b  &  1821  &  1.7  &  2.99  &  0.69  &  8.69  &  20.5  &  21.5  &  29.7 \\
 &  WASP-19 b  &  2117  &  1.42  &  0.79  &  0.67  &  10.48  &  20.2  &  22.6  &  30.3 \\
 &  HAT-P-70 b  &  2552  &  1.87  &  2.74  &  0.62  &  8.96  &  18.9  &  24.0  &  30.5 \\
 &  KELT-7 b  &  2042  &  1.6  &  2.73  &  0.6  &  7.54  &  18.4  &  24.7  &  30.8 \\
 &  WASP-43 b  &  1379  &  0.93  &  0.81  &  0.66  &  9.27  &  20.2  &  23.3  &  30.8 \\
 &  KELT-8 b  &  1676  &  1.62  &  3.24  &  0.75  &  9.18  &  23.2  &  20.3  &  30.8 \\
 &  TOI-2109 b  &  3072  &  1.35  &  0.67  &  0.75  &  9.07  &  23.5  &  20.9  &  31.5 \\
 &  KELT-14 b  &  1961  &  1.74  &  1.71  &  0.86  &  9.42  &  27.4  &  16.1  &  31.7 \\
 &  WASP-74 b  &  1916  &  1.36  &  2.14  &  0.86  &  8.22  &  27.8  &  16.2  &  32.2 \\
 &  KELT-23 A b  &  1565  &  1.32  &  2.26  &  0.57  &  8.9  &  18.9  &  27.0  &  32.9 \\
 &  HD 209458 b  &  1453  &  1.39  &  3.52  &  0.5  &  6.31  &  16.7  &  28.7  &  33.2 \\
 &  KELT-19 A b  &  1938  &  1.91  &  4.61  &  0.6  &  9.2  &  20.4  &  27.3  &  34.1 \\
 &  TOI-1431 b  &  2395  &  1.49  &  2.65  &  0.88  &  7.44  &  30.2  &  16.0  &  34.2 \\
    \end{tabular}
    \caption{The best 15 targets for eclipse mapping with NIRSpec G395H, ordered by their single-eclipse overall resolution EMM according to Equation \ref{eq:EMM}. HD 189733 b is marked with an asterisk as it is expected to partially saturate the detector but we leave it here for reference.}
    \label{tab:nirspec_targets}
\end{table*}

\begin{table*}
    \centering
    \begin{tabular}{rl|cccccccc}
\multicolumn{2}{c}{\textbf{Planet}}  &  \textbf{Equilibrium}  &  \textbf{Radius}  &  \textbf{Period} & \textbf{Impact}  & \textbf{K}  &  \textbf{EMM}$_{x}$  &  \textbf{EMM}$_{y}$ &  \textbf{EMM} \\
\multicolumn{2}{c}{\textbf{Name}}  &  \textbf{Temperature (K)}  &  ($R_{J}$)  &  \textbf{(days)} & \textbf{Parameter}& \textbf{(mag)} & \textbf{(deg)} &  \textbf{(deg)}  &  \textbf{(deg)} \\
\hline
&  HD 189733 b  &  1202  &  1.13  &  2.22  &  0.66  &  5.54  &  15.3  &  17.4  &  23.2 \\
 &  HD 209458 b  &  1453  &  1.39  &  3.52  &  0.5  &  6.31  &  15.7  &  26.9  &  31.2 \\
 &  KELT-20 b  &  2255  &  1.74  &  3.47  &  0.51  &  7.42  &  16.7  &  28.4  &  33.0 \\
 &  WASP-69 b  &  960  &  1.11  &  3.87  &  0.69  &  7.46  &  22.8  &  24.1  &  33.2 \\
 &  WASP-43 b  &  1379  &  0.93  &  0.81  &  0.66  &  9.27  &  22.3  &  25.7  &  34.0 \\
 &  KELT-4 A b  &  1821  &  1.7  &  2.99  &  0.69  &  8.69  &  23.5  &  24.6  &  34.1 \\
 &  KELT-7 b  &  2042  &  1.6  &  2.73  &  0.6  &  7.54  &  20.4  &  27.5  &  34.3 \\
 &  KELT-8 b  &  1676  &  1.62  &  3.24  &  0.75  &  9.18  &  26.8  &  23.4  &  35.6 \\
 &  WASP-74 b  &  1916  &  1.36  &  2.14  &  0.86  &  8.22  &  31.6  &  18.4  &  36.5 \\
 &  KELT-23 A b  &  1565  &  1.32  &  2.26  &  0.57  &  8.9  &  21.1  &  30.2  &  36.9 \\
 &  HAT-P-70 b  &  2552  &  1.87  &  2.74  &  0.62  &  8.96  &  23.1  &  29.2  &  37.2 \\
 &  WASP-189 b  &  2636  &  1.62  &  2.72  &  0.48  &  6.06  &  18.2  &  33.4  &  38.1 \\
 &  KELT-14 b  &  1961  &  1.74  &  1.71  &  0.86  &  9.42  &  33.2  &  19.5  &  38.5 \\
 &  TOI-1431 b  &  2395  &  1.49  &  2.65  &  0.88  &  7.44  &  34.0  &  18.1  &  38.5 \\
 &  TOI-778 b  &  1710  &  1.37  &  4.63  &  0.7  &  8.06  &  27.1  &  27.9  &  38.9 \\
    \end{tabular}
    \caption{The best 15 targets for eclipse mapping with MIRI LRS, ordered by their single-eclipse overall resolution EMM according to Equation \ref{eq:EMM}.}
    \label{tab:miri_targets}
\end{table*}

\section{Conclusions}\label{sec:conclusions}
We derived an analytic theory for the resolution achievable using eclipse mapping on an exoplanet. We find that:
\begin{enumerate}
    \item The space of surface brightness patterns is highly degenerate in the space of eclipse light-curves; that is, there are many brightness patterns which all result in the same light-curve. Equivalently, there are very few brightness patterns which yield non-zero light-curves compared to the number of brightness patterns whose light-curves are zero (the "null space"; see \citet{challener2023eclipse}). Nonetheless, there are particular brightness patterns which contribute high-frequency components not found in the light-curves of other patterns at the same wavenumber, and these can then be constrained much more easily.
    \item The signal from a mode of a given wavenumber and amplitude scales as the inverse of the wavenumber; phase curve inversion scales as the inverse square of the wavenumber \citep{cowan2008inverting} so eclipse mapping is able to resolve smaller details for the same amount of observing time.
    \item Brightness patterns whose features align with the edge of the star during ingress and egress yield large, high-frequency signals, while others experience cancellation between bright and dark regions and thus yield small signals. As such, we can constrain the well-aligned brightness patterns well, and have to use other methods (e.g. imposing non-negativity or maximising smoothness) to constrain the others. Because of this, resolutions in the latitude and longitude directions are affected strongly by the impact parameter: at high impact parameters, the longitudinal resolution is worsened and small features are not detectable, while the latitudinal resolution is good. Thus the choice of target planet depends strongly on what features one intends to observe: a planet with high impact parameter is ideal for measuring the width of the equatorial jet, while for measuring steep of east-west brightness temperature gradients (e.g. where cloud formation starts), one should opt for a planet with an impact parameter near to zero.
    \item We derive an Eclipse Mapping Metric (Equation \ref{eq:EMM}) which estimates the achievable resolution in latitude and longitude; we calculate this for the MIRI LRS, NIRSpec G395H, and NIRISS SOSS instruments on JWST. We find our values to be consistent with JWST observations of WASP-18 b \citep{coulombe2023broadband} which constrain well features along the longitude axis, but where features in latitude were almost unconstrained.
\end{enumerate}
We find our theoretical predictions validated by our numerical tests, which yield the correct resolutions with an error of 10\% from the predicted values, as well as the correct scaling (to a similar accuracy) for the number of resolvable modes as a function of light curve error. Given our theory, we present a list of exoplanets with particularly good resolution for eclipse mapping.

\section*{Acknowledgements}

S.B. gratefully acknowledges funding from the Research Centre at Christ Church, Oxford. M.H. gratefully acknowledges funding from Christ Church, Oxford. D.G. acknowledges funding from the UKRI STFC Consolidated Grant ST/V000454/1.

\section*{Data Availability}

The data underlying this article are available in the article and in its online supplementary material. The code used to generate the data is available on request.


\bibliographystyle{mnras}
\bibliography{main} 

\begin{thebibliography}{}
\makeatletter
\relax
\def\mn@urlcharsother{\let\do\@makeother \do\$\do\&\do\#\do\^\do\_\do\%\do\~}
\def\mn@doi{\begingroup\mn@urlcharsother \@ifnextchar [ {\mn@doi@}
  {\mn@doi@[]}}
\def\mn@doi@[#1]#2{\def\@tempa{#1}\ifx\@tempa\@empty \href
  {http://dx.doi.org/#2} {doi:#2}\else \href {http://dx.doi.org/#2} {#1}\fi
  \endgroup}
\def\mn@eprint#1#2{\mn@eprint@#1:#2::\@nil}
\def\mn@eprint@arXiv#1{\href {http://arxiv.org/abs/#1} {{\tt arXiv:#1}}}
\def\mn@eprint@dblp#1{\href {http://dblp.uni-trier.de/rec/bibtex/#1.xml}
  {dblp:#1}}
\def\mn@eprint@#1:#2:#3:#4\@nil{\def\@tempa {#1}\def\@tempb {#2}\def\@tempc
  {#3}\ifx \@tempc \@empty \let \@tempc \@tempb \let \@tempb \@tempa \fi \ifx
  \@tempb \@empty \def\@tempb {arXiv}\fi \@ifundefined
  {mn@eprint@\@tempb}{\@tempb:\@tempc}{\expandafter \expandafter \csname
  mn@eprint@\@tempb\endcsname \expandafter{\@tempc}}}

\bibitem[\protect\citeauthoryear{Adams \& Rauscher}{Adams \&
  Rauscher}{2022}]{adams2022sensitivity}
Adams A.~D.,  Rauscher E.,  2022, The Astronomical Journal, 165, 24

\bibitem[\protect\citeauthoryear{Agol, Cowan, Knutson, Deming, Steffen, Henry
  \& Charbonneau}{Agol et~al.}{2010}]{agol2010climate}
Agol E.,  Cowan N.~B.,  Knutson H.~A.,  Deming D.,  Steffen J.~H.,  Henry
  G.~W.,   Charbonneau D.,  2010, The Astrophysical Journal, 721, 1861

\bibitem[\protect\citeauthoryear{Baptista \& Bortoletto}{Baptista \&
  Bortoletto}{2004}]{baptista2004eclipse}
Baptista R.,  Bortoletto A.,  2004, The Astronomical Journal, 128, 411

\bibitem[\protect\citeauthoryear{Batalha et~al.,}{Batalha
  et~al.}{2017}]{batalha2017pandexo}
Batalha N.~E.,  et~al., 2017, Publications of the Astronomical Society of the
  Pacific, 129, 064501

\bibitem[\protect\citeauthoryear{Challener \& Rauscher}{Challener \&
  Rauscher}{2023}]{challener2023eclipse}
Challener R.~C.,  Rauscher E.,  2023, arXiv preprint arXiv:2309.05539

\bibitem[\protect\citeauthoryear{Cho, Menou, Hansen  \& Seager}{Cho
  et~al.}{2003}]{cho2003changing}
Cho J.~Y.,  Menou K.,  Hansen B.~M.,   Seager S.,  2003, The Astrophysical
  Journal, 587, L117

\bibitem[\protect\citeauthoryear{Collier~Cameron}{Collier~Cameron}{1997}]{collier1997eclipse}
Collier~Cameron A.,  1997, Monthly Notices of the Royal Astronomical Society,
  Volume 287, Issue 3, pp. 556-566., 287, 556

\bibitem[\protect\citeauthoryear{Coulombe et~al.,}{Coulombe
  et~al.}{2023}]{coulombe2023broadband}
Coulombe L.-P.,  et~al., 2023, Nature, pp~1--3

\bibitem[\protect\citeauthoryear{Cowan \& Agol}{Cowan \&
  Agol}{2008}]{cowan2008inverting}
Cowan N.~B.,  Agol E.,  2008, The Astrophysical Journal, 678, L129

\bibitem[\protect\citeauthoryear{Cowan \& Fujii}{Cowan \&
  Fujii}{2017}]{cowan2017mapping}
Cowan N.~B.,  Fujii Y.,  2017, arXiv preprint arXiv:1704.07832

\bibitem[\protect\citeauthoryear{Cowan, Fuentes  \& Haggard}{Cowan
  et~al.}{2013}]{cowan2013light}
Cowan N.~B.,  Fuentes P.~A.,   Haggard H.~M.,  2013, Monthly Notices of the
  Royal Astronomical Society, 434, 2465

\bibitem[\protect\citeauthoryear{Dobbs-Dixon, Agol  \& Deming}{Dobbs-Dixon
  et~al.}{2015}]{dobbs2015spectral}
Dobbs-Dixon I.,  Agol E.,   Deming D.,  2015, The Astrophysical Journal, 815,
  60

\bibitem[\protect\citeauthoryear{Guillot, Burrows, Hubbard, Lunine  \&
  Saumon}{Guillot et~al.}{1996}]{guillot1996giant}
Guillot T.,  Burrows A.,  Hubbard W.,  Lunine J.,   Saumon D.,  1996, The
  Astrophysical Journal, 459, L35

\bibitem[\protect\citeauthoryear{Horne}{Horne}{1985}]{horne1985images}
Horne K.,  1985, Monthly Notices of the Royal Astronomical Society, 213, 129

\bibitem[\protect\citeauthoryear{Jackson, Adams, Sandidge, Kreyche  \&
  Briggs}{Jackson et~al.}{2019}]{jackson2019variability}
Jackson B.,  Adams E.,  Sandidge W.,  Kreyche S.,   Briggs J.,  2019, The
  Astronomical Journal, 157, 239

\bibitem[\protect\citeauthoryear{Knutson et~al.,}{Knutson
  et~al.}{2007}]{knutson2007map}
Knutson H.~A.,  et~al., 2007, Nature, 447, 183

\bibitem[\protect\citeauthoryear{Lewis \& Hammond}{Lewis \&
  Hammond}{2022}]{lewis2022temperature}
Lewis N.~T.,  Hammond M.,  2022, The Astrophysical Journal, 941, 171

\bibitem[\protect\citeauthoryear{Louden \& Kreidberg}{Louden \&
  Kreidberg}{2018}]{louden2018spiderman}
Louden T.,  Kreidberg L.,  2018, Monthly Notices of the Royal Astronomical
  Society, 477, 2613

\bibitem[\protect\citeauthoryear{Luger, Agol, Foreman-Mackey, Fleming,
  Lustig-Yaeger  \& Deitrick}{Luger et~al.}{2019}]{luger2019starry}
Luger R.,  Agol E.,  Foreman-Mackey D.,  Fleming D.~P.,  Lustig-Yaeger J.,
  Deitrick R.,  2019, The astronomical journal, 157, 64

\bibitem[\protect\citeauthoryear{Majeau, Agol  \& Cowan}{Majeau
  et~al.}{2012}]{majeau2012two}
Majeau C.,  Agol E.,   Cowan N.~B.,  2012, The Astrophysical Journal Letters,
  747, L20

\bibitem[\protect\citeauthoryear{Menou}{Menou}{2012}]{menou2012magnetic}
Menou K.,  2012, The Astrophysical Journal, 745, 138

\bibitem[\protect\citeauthoryear{Menou}{Menou}{2020}]{menou2020hot}
Menou K.,  2020, Monthly Notices of the Royal Astronomical Society, 493, 5038

\bibitem[\protect\citeauthoryear{Mullally, Rodriguez, Stevenson  \&
  Wakeford}{Mullally et~al.}{2019}]{mullally2019exo}
Mullally S.~E.,  Rodriguez D.~R.,  Stevenson K.~B.,   Wakeford H.~R.,  2019,
  Res. Notes AAS, 3, 193

\bibitem[\protect\citeauthoryear{Parmentier \& Crossfield}{Parmentier \&
  Crossfield}{2017}]{parmentier2017exoplanet}
Parmentier V.,  Crossfield I.,  2017, arXiv preprint arXiv:1711.07696

\bibitem[\protect\citeauthoryear{Parmentier, Showman  \& Lian}{Parmentier
  et~al.}{2013}]{parmentier20133d}
Parmentier V.,  Showman A.~P.,   Lian Y.,  2013, Astronomy \& Astrophysics,
  558, A91

\bibitem[\protect\citeauthoryear{Parmentier, Fortney, Showman, Morley  \&
  Marley}{Parmentier et~al.}{2016}]{parmentier2016transitions}
Parmentier V.,  Fortney J.~J.,  Showman A.~P.,  Morley C.,   Marley M.~S.,
  2016, The Astrophysical Journal, 828, 22

\bibitem[\protect\citeauthoryear{Parmentier et~al.,}{Parmentier
  et~al.}{2018}]{parmentier2018thermal}
Parmentier V.,  et~al., 2018, Astronomy \& Astrophysics, 617, A110

\bibitem[\protect\citeauthoryear{Perez-Becker \& Showman}{Perez-Becker \&
  Showman}{2013}]{perez2013atmospheric}
Perez-Becker D.,  Showman A.~P.,  2013, The Astrophysical Journal, 776, 134

\bibitem[\protect\citeauthoryear{Rasio, Tout, Lubow  \& Livio}{Rasio
  et~al.}{1996}]{rasio1996tidal}
Rasio F.,  Tout C.,  Lubow S.,   Livio M.,  1996, arXiv preprint
  astro-ph/9605059

\bibitem[\protect\citeauthoryear{Rauscher, Menou, Cho, Seager  \&
  Hansen}{Rauscher et~al.}{2007a}]{rauscher2007hot}
Rauscher E.,  Menou K.,  Cho J. Y.-K.,  Seager S.,   Hansen B.~M.,  2007a, The
  Astrophysical Journal, 662, L115

\bibitem[\protect\citeauthoryear{Rauscher, Menou, Seager, Deming, Cho  \&
  Hansen}{Rauscher et~al.}{2007b}]{rauscher2007toward}
Rauscher E.,  Menou K.,  Seager S.,  Deming D.,  Cho J. Y.-K.,   Hansen B.~M.,
  2007b, The Astrophysical Journal, 664, 1199

\bibitem[\protect\citeauthoryear{Roman, Kempton, Rauscher, Harada, Bean  \&
  Stevenson}{Roman et~al.}{2021}]{roman2021clouds}
Roman M.~T.,  Kempton E. M.-R.,  Rauscher E.,  Harada C.~K.,  Bean J.~L.,
  Stevenson K.~B.,  2021, The Astrophysical Journal, 908, 101

\bibitem[\protect\citeauthoryear{Showman \& Guillot}{Showman \&
  Guillot}{2002}]{showman2002atmospheric}
Showman A.~P.,  Guillot T.,  2002, Astronomy \& Astrophysics, 385, 166

\bibitem[\protect\citeauthoryear{Showman, Tan  \& Parmentier}{Showman
  et~al.}{2020}]{showman2020atmospheric}
Showman A.~P.,  Tan X.,   Parmentier V.,  2020, Space Science Reviews, 216, 1

\bibitem[\protect\citeauthoryear{Williams, Charbonneau, Cooper, Showman  \&
  Fortney}{Williams et~al.}{2006}]{williams2006resolving}
Williams P.~K.,  Charbonneau D.,  Cooper C.~S.,  Showman A.~P.,   Fortney
  J.~J.,  2006, The Astrophysical Journal, 649, 1020

\bibitem[\protect\citeauthoryear{de Wit, Gillon, Demory  \& Seager}{de~Wit
  et~al.}{2012}]{de2012towards}
de Wit J.,  Gillon M.,  Demory B.-O.,   Seager S.,  2012, Astronomy \&
  Astrophysics, 548, A128

\makeatother
\end{thebibliography}






\bsp	
\label{lastpage}
\end{document}